\documentclass[11pt]{article}

\usepackage{a4}
\usepackage{times}
\usepackage{graphicx}
\usepackage{amsmath}
\usepackage{amsfonts}
\usepackage{amssymb}
\usepackage{natbib}

\newcommand{\nats}{{\mathbb N}}

\newcommand{\D}{{\mathcal D}}
\newcommand{\E}{{\mathcal E}}
\newcommand{\G}{{\mathcal G}}
\newcommand{\K}{{\mathcal K}}
\newcommand{\N}{{\mathcal N}}
\newcommand{\X}{{\mathcal X}}

\newcommand{\var}{\operatorname{var}}
\newcommand{\ad}{\operatorname{ad}}
\newcommand{\cov}{\operatorname{cov}}

\newcommand{\bZ}{{\mathbf Z}}
\newcommand{\bY}{{\mathbf Y}}
\newcommand{\bX}{{\mathbf X}}
\newcommand{\bW}{{\mathbf W}}
\newcommand{\bepsilon}{{\boldsymbol \epsilon}}

\newcommand{\Dfrak}{{\mathfrak D}}

\title{Modelling, Detrending and Decorrelation of Network Time Series}
\author{Marina Knight, Matt Nunes and Guy Nason\thanks{Corresponding author: Guy P.\ Nason, School of Mathematics, University of Bristol, Bristol UK; g.p.nason@bristol.ac.uk}}

\date{29th February 2016} 

\begin{document}
\maketitle

\begin{abstract}
A network time series is a multivariate time series augmented by a graph
that describes how variables (or nodes) are connected. We introduce
the network autoregressive (integrated) moving average (NARIMA)
processes: a set
of flexible models for network time series. For fixed networks the NARIMA
models are essentially equivalent to vector autoregressive moving average-type
models. However, NARIMA models are especially useful when the  structure
of the graph, associated with the multivariate time series, changes over time.
Such network topology changes are invisible to standard VARMA-like
models. For integrated NARIMA models
we introduce network differencing,
based on the network lifting (wavelet) transform, which removes trend. We exhibit our techniques
on a network time series describing the evolution of mumps
throughout counties of England and Wales weekly during 2005. We further
demonstrate the action of network lifting on a simple bivariate VAR(1) model
with associated two-node graph. We show theoretically that decorrelation
occurs only in certain circumstances and maybe less 
than expected. This suggests that the time-decorrelation properties of spatial
network lifting are due more to the trend removal properties of lifting
rather than any kind of stochastic decorrelation.
\end{abstract}

\section{Background}
Recently, the analysis of data on graphs through time (network time series) has become of increasing
importance. We are
now able to collect not only large multivariate time series but also strong and useful information
on how individuals (variables) in those multivariate time series are related to each other
via a graph (network) description.
Such hard network information is typically
more powerful than just relying on measures of association (correlation)
between variables to gauge their real relationship. We are primarily interested in stochastic processes
observed at nodes of a graph over time and so we use `variable' and `node' interchangeably as context
demands.

Suitable models for multivariate (or vector) time series have existed for a long time.
For example, vector autoregression (VAR) models,
see \cite{hamilton:time}, Chapter~11 or \cite{Wei06}, Chapter~16.
This article proposes a new class of models for network time series:
the network autoregressive (integrated) moving average (NARIMA) processes.  NARIMA models borrow
heavily
from existing multivariate time series models and, for fixed networks in their simplest form, they are
statistically equivalent to a VAR.
However, the NARIMA structure forces analysts to undertake
a different approach to data modelling as the network structure influences the type of NARIMA (or VAR)
model fitted.  VAR model fitting requires various  approaches  to be undertaken to reduce
the model dimension to a manageable size. In NARIMA models such modelling is considerably
aided by the network structure.

A major advantage of NARIMA models is how they easily cope with the ``moving node'' effect in
dynamic networks.
Many networks are not static. In particular, nodes can change their position within the network 
structure or they can disappear or reappear. A good example of this effect often
arises in epidemiological studies. For example, in the evolution of a foot and mouth disease outbreak
infected individuals (cows) can die (or be born) but crucially they can move around the network as they
get bought, or sold or taken to market or quarantined. A standard multivariate time series model (like VAR)
only looks at the number of cases in a herd (for example) and has no explicit mechanism for incorporating
the ``moving node'' effect. However, NARIMA models can
explicitly track this important and valuable information
and use it for modelling and forecasting purposes.

Another interesting, and hard to deal with,
 feature of network time series is that of trend. Trend is well-handled in regular
time series by the important techniques such as differencing or fitting curves and, as such,
trend
removal is a vital component of time series, see \cite{chatfield:the}.
Unremoved  trend can severely
distort estimation of remaining stochastic structure. We propose using the network lifting method
from \cite{JansenNasonSilverman01} and \cite{JNS09} to estimate and remove trend from 
networks as a preprocessing step that can be used prior to modelling using  NARIMA models.
Finally, we consider a simple prototype NARIMA model and show theoretically, for the first time,
 the conditions under which ({\em spatial}) network lifting can achieve decorrelation in {\em time}.

\section{Setup and Notation}
This article is concerned with data collected on graphs (networks).
Our graph, $\G = (\K, \E)$,
consists of a set of nodes, $\K$, some of which are joined together by edges from some set of
edges $\E$.
We define the set of $K$ nodes $\K = \{ 1, \ldots, K\}$.
Two nodes, $i, j \in \K$, connected by an (undirected) edge are denoted by $i \leftrightsquigarrow j$.
The set of edges in the graph are defined by
$\E  = \{ (i, j) : i \leftrightsquigarrow j; i, j \in \K \}$.
Sometimes, the set of edges is supplemented by another set, the edge distances
$\D$ which merely contains the distance $d(i,j)$ between nodes $i, j$ when
$i \leftrightsquigarrow j$.

Suppose $A \subset \K$ is a subset of nodes. The neighbourhood set of $A$ is defined by
$\N (A ) = \{ j \in \K/ A: j \leftrightsquigarrow i, i\in A \}$. We define the set of
$r$th-stage neighbours of a node $i \in \K$ by
\begin{equation}
\N^{(r)} (i) = \N \{ \N^{(r-1)} (i) \} / \cup_{q=1}^{(r-1)} \N^{(q)} (i),
\end{equation}
for $r= 2, 3, \ldots$ and $\N^{(1)} (i) =  \N (\{ i \})$. In other words, $\N^{(r)}(i)$ is the
set of any points connected to any element of $\N^{(r-1)}(i)$ by an edge that has not appeared
in any earlier neighbourhood set.

Initially, we will consider functions that are evaluated at the nodes, and further interested in these
values as functions of time. So, we consider $T$ time points, $t_1, \ldots, t_T$, initially focusing
on $t_m = m \in \nats$. A key part  of a
network time series is its component multivariate time series $X_{i, t}$ 
for the value of the time series at node $i \in \K(\G)$ at time $t \in 1, \ldots, T$,
where $\K (\G)$ is the set of nodes associated with graph $\G$. Then
\begin{center}
\framebox{A network time series is $\X = ( \{ X_{i, t} \}_{i\in \K(\G), t=1}^{T}, \G)$.}
\end{center}

\begin{figure}
\centering
\resizebox{\textwidth}{!}{\includegraphics{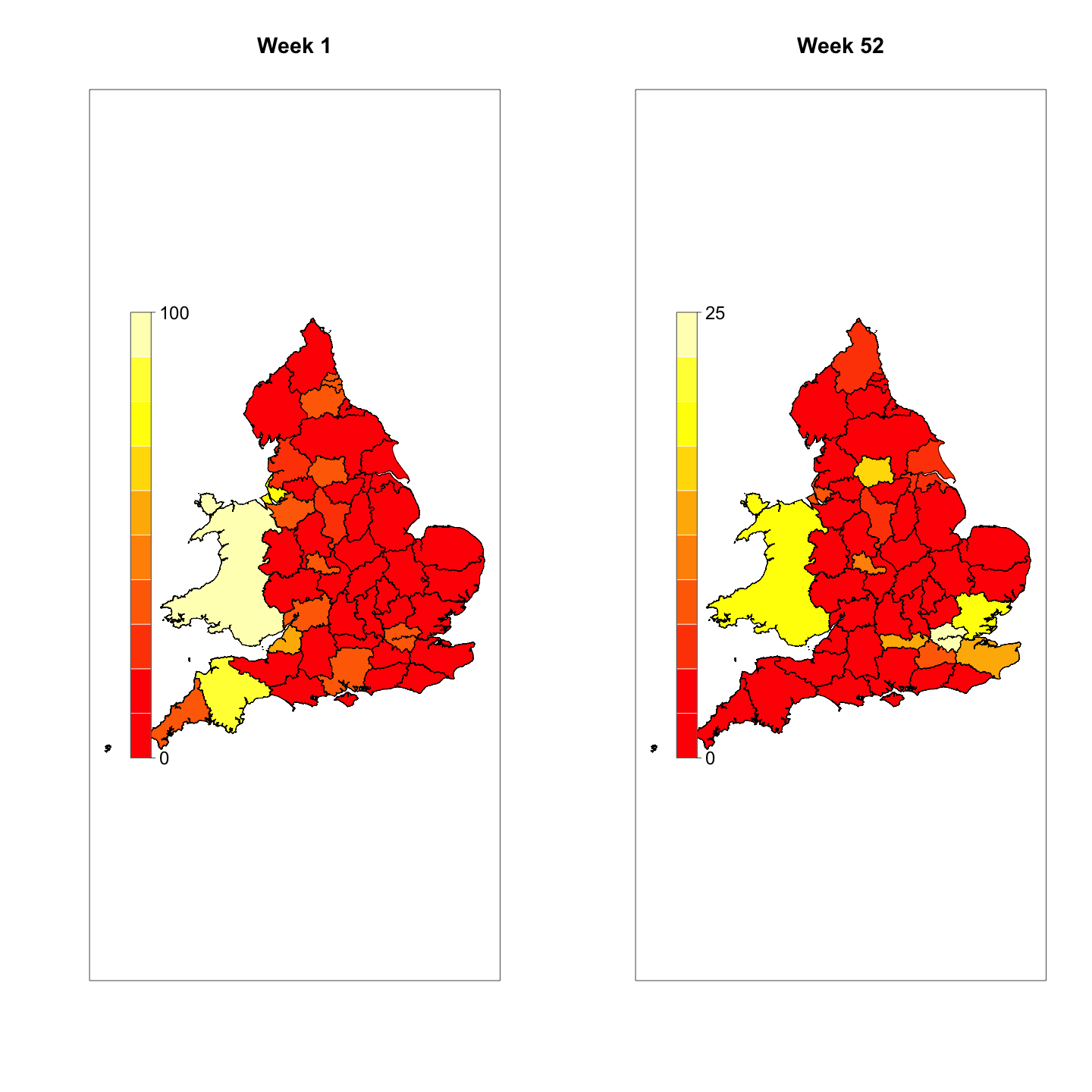}}
\caption{Number of cases of Mumps in each UK county during week 1 (left) and week 52 (right).
\label{fig:mumpsukpic}}
\end{figure}

{\em Mumps example:}
We study  the number of cases of mumps in each
of 47 counties of 
England and Wales taken weekly during 2005 from week one to week 52.
This period of time was particularly interesting for this disease as it was
shortly after the MMR scare which resulted in an abnormally large proportion
of the relevant population  not receiving the MMR vaccine.
Figure~\ref{fig:mumpsukpic} shows the situation for the first and last week of 2005
with number of cases colour-coded with deep red indicating few cases and yellow
through to white indicating a large number of cases. For example, during week one
Wales (which is treated as a single county here) and Devon have very high counts and
in week 52 Wales is still high, as is Essex, but Devon's cases have subsided.
In this example, $X_{i, t}$ is the multivariate time series of the number of cases of mumps
in county $i=1, \ldots, 47$ for weeks $t=1, \ldots, 52$.

\begin{figure}
\centering
\resizebox{\textwidth}{!}{\includegraphics{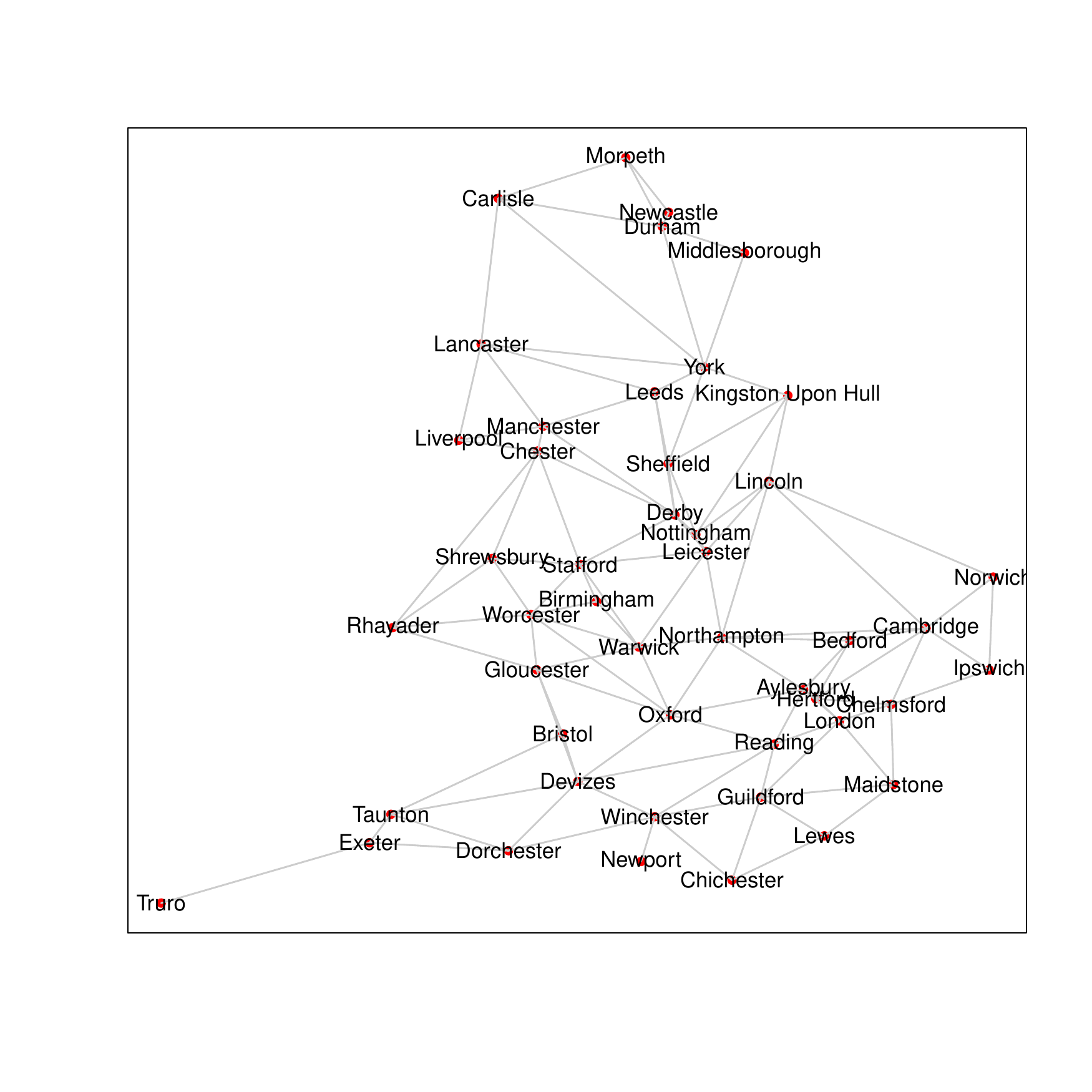}}
\caption{Graph showing connected edges between UK county towns.
\label{fig:mumpsuknet}}
\end{figure}

Figure~\ref{fig:mumpsuknet} shows the graph, $\G$, associated with our network
time series. This particular graph was constructed by identifying a ``county town''
for each county (and Rhayader for Wales) and the constructing a graph that connects
all towns less than radius of a predefined fixed number of kilometres
which reflects the strength of communication
links between different parts of the UK. Of course, depending on the disease
epidemiology, different graphs could be constructed. For example, in animal
diseases such as foot and mouth routes between farms and between farms
and markets, as well as geographical proximity to allow for spread of the
disease vector by wind would be instrumental in the development of a suitable graph.

We will now introduce some models for network time series.
\section{Network Autoregressive Moving Average Model (NARMA)}
Suppose that $\X$ is a network time series. A {\em network autoregressive process of order $p$ and
neighbourhood order vector $s$ of length $p$}, denoted \mbox{NAR$(p, s)$}, is given by:
\begin{equation}
\label{eq:narps}
X_{i, t} = \sum_{j=1}^p \left(  \alpha_j X_{i, t-j} + \sum_{r=1}^{s_j} \sum_{q \in \N^{(r)}(i)} \beta_{j, r, q} X_{q, t-j} 
	\right) 
	+ \epsilon_{i, t},
\end{equation}
where, for this article at least, we assume $\epsilon_{i, t}$ are a set of mutually uncorrelated
random variables with mean zero and variance of $\sigma^2$.

Writing the vector $\bX_t = (X_{1, t}, \ldots, X_{K, t} )^T$ and letting
$\{\bZ_t \}$ be a standard vector moving average model of order $q$,
with first term $(\epsilon_{1,t }, \ldots, \epsilon_{K, t})$, a {\em network
autoregressive moving average process} of order $(p, s; q)$ is given by $\bY_t = \bX_t + \bZ_t$.
Clearly, such a model specification can get quite complicated, particularly for larger $p, q$ and the 
majority of  this article deals with NAR or NARI models.

An integrated model can be obtained after some differencing operator, $\Dfrak$, is
applied to some network $\bX_t$ by  $\bW_t = \Dfrak \bX_t$  and $\bW_t$
is then  modelled as a NARMA process. This is analogous to the
standard ARIMA mode of operations for univariate time series. We shall say more on
`network differencing' or detrending in Section~\ref{sec:detrend}.

Model~\eqref{eq:narps} expresses how past values of the network time series influence the current values.
In particular, $X_{t, i}$ depends directly on its past values at that node via the $\alpha_j$ term and
also on past values of its neighbours (and neighbours of neighbours, etc) through the $\beta_{j, r, q}$ term.
Our NAR$(p, s)$ model assumes that the $\{ \alpha_j \}$ and $\{ \beta_{j, r, q} \}$ parameter
sets do not depend on $t$ (hence, stationarity is assumed), neither do they depend on the node $i$ 
(spatial homogeneity). Naturally, both of these assumptions might be questioned in real examples and
the model extended.

A NAR$\{ p, (0, 0, \ldots, 0) \}$ means a model consisting of
$K$ separate regular AR$(p)$ time series models, one for each node.
A general NAR$(p, s)$ model, as it is shown in~\eqref{eq:narps} can be
viewed as a vector autoregressive VAR model with a specific set of constraints
on the VAR parameters. However, the modelling process is somewhat different
with NARIMA models as past regressors depend on neighbours (and stage-$r$
neighbours) of all nodes in the same way.
More importantly, the way in which node neighbour contributions are constructed
in NARIMA models is specialized and strongly relate to the structure and topology
of the associated graph $\G$.

Later, though, we will define
a gNARIMA process which is similar to NARIMA except that nodes can drop-out and
reappear arbitrarily which is not covered by VAR. One might think that
multivariate time series with missing observations can deal with this case, but
nodes can disappear and reappear and change their geometry within the network
in the meantime. So, in terms of the node and what it represents it refers to the
same object, but its position in the graph might be quite different. gNARIMA models
can handle this, whereas VAR can not.

\subsection{NAR$(1, 1)$ example}
To explain the key features of network autoregressive models we focus on the NAR$(p, s)$ model
with $p=1$ and $s=1$ which can be written as
\begin{equation}
\label{eq:nar11}
X_{i, t} = \alpha X_{i, t-1} + \sum_{q \in \N^{(1)} (i)} \beta_q X_{q, t-1} + \epsilon_{i, t},
\end{equation}
here we can drop the $j$ and $r$ subscripts for a simplified presentation.
In this simpler example the value at node $X_{i, t}$ depends directly on
$X_{i, t-1}$ in the usual autoregressive way, but also depends on the
neighbours of node $i$ at the previous time step $t-1$.

There are several modelling choices to be made for network autoregressive models.
We might choose to incorporate distance information into
our specification of $\beta_q$ such as weighting neighbours of $i$ more
if they are closer to $i$. For example, we might compute inverse distance
weights $w_j (i) = d(i, j)/ \sum_{k \in \N(i)} d(i, k)$ for $j \in \N(i)$.
Then we might parametrise $\beta_q$ in~\eqref{eq:nar11} by
\begin{equation}
\label{eq:idbetaspec}
\beta_q = \beta \, w_q (i),
\end{equation}
for $q \in \N (i)$. This model specifies the overall first-stage neighbour autoregression
strength by $\beta$ but modulated by the inverse distance weights.
With a gNARIMA model we permit the weights to change as a node changes its position
(or existence) within the network topology. Such changes can be easily incorporated into
the least-squares estimation process as the overall model description does not change.

\subsection{NAR$(1,1)$ modelling for mumps data}
We can use the {\tt nar()} function in R to model mumps using the NAR$(1,1)$
model as an example and examine the fit. We first apply the modelling
to the disease incidence, that is we divide the raw mumps counts by an estimate
of the population size for each county. This information is stored in the
network time series {\tt mumpsPcor}.

\begin{figure}
\centering
\resizebox{\textwidth}{!}{\includegraphics{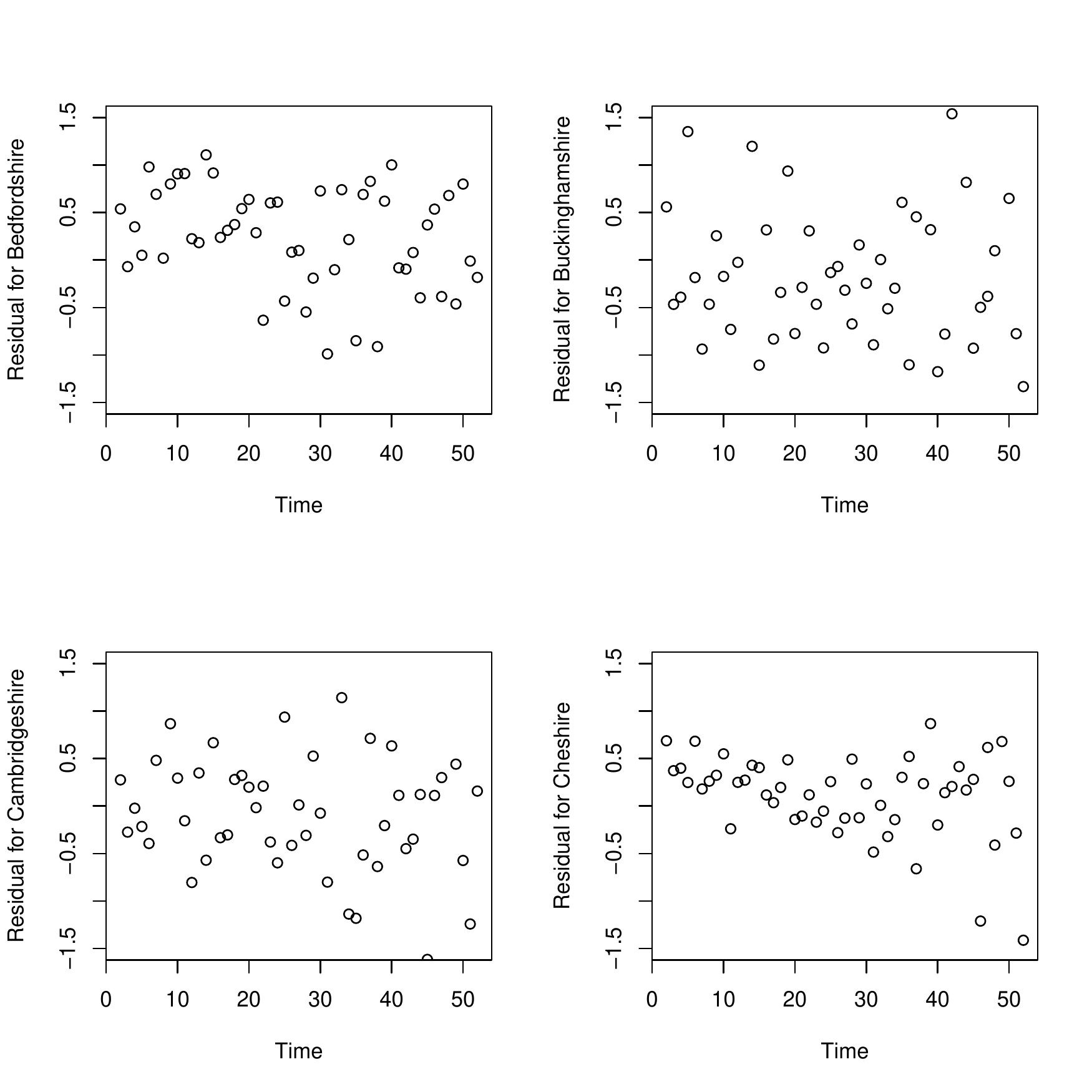}}
\caption{Residual plots for Bedfordshire, Buckinghamshire, Cambridgeshire
	and Cheshire after fitting NAR$(1,1)$ model with inverse distance
	weight $\beta$ specification on logarithmic population-corrected mumps counts.\label{fig:model2res}}
\end{figure}

We modelled the {\tt mumpsPcor} series using the NAR$(1,1)$ model with
the inverse distance weights specification for $\beta_q$ given in~\eqref{eq:idbetaspec}
using the command:
\begin{verbatim}
model1 <- nar(vts=mumpsPcor, net=townnet2)
\end{verbatim}
The model is fitted using least squares although it is easy to see that
maximum likelihood or Bayesian inference might well be preferable in some
circumstances, particularly when it comes to formulating uncertainty measures.
From this model fit we estimated $\hat{\alpha} \approx 0.683$ and
$\hat{\beta} \approx 0.263$. We also compute the residual matrix:
\begin{equation}
r_{i, t} = \hat{\epsilon}_{i, t} = X_{i, t} - \hat{\alpha} X_{i, t-1}
	- \sum_{q \in \N^{(1)} (i)} \hat{\beta}_q X_{q, t-1},
\end{equation}
for $t=2, \ldots, 52$.
The residuals are, as usual, vital for assessment of model fit.
\begin{figure}
\centering
\resizebox{\textwidth}{!}{\includegraphics{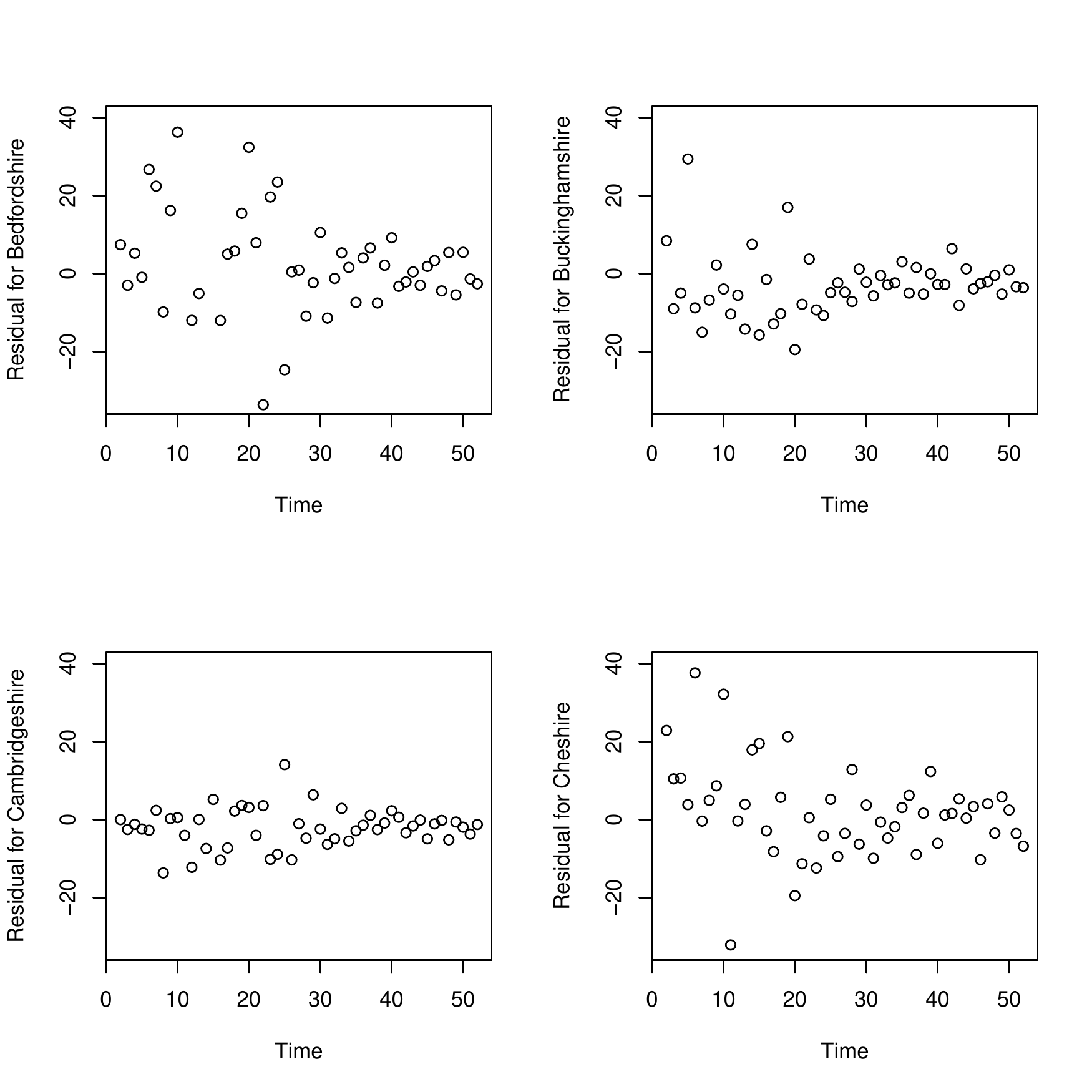}}
\caption{Residual plots for Bedfordshire, Buckinghamshire, Cambridgeshire
	and Cheshire after fitting NAR$(1,1)$ model with inverse distance
	weight $\beta$ specification on population-corrected mumps counts.\label{fig:model1res}}
\end{figure}
Figure~\ref{fig:model1res} shows the residuals plotted against time for
four counties and one can see that the residuals indicate that the
model is not a good fit as the variance appears not to be constant over time.
In time-honoured tradtion we apply a variance stabilizing logarithmic transform.
More precisely, we model $Y_{i, t} = \log ( 1 + X_{i, t})$ and this
new multivariate time series is stored in the object {\tt LmumpsPcor}.
We refit the new model by
\begin{verbatim}
model2 <- nar(vts=LmumpsPcor, net=townnet2)
\end{verbatim}
which results in new estimates of $\hat{\alpha} \approx 0.647$
and $\hat{\beta} \approx 0.330$.
Similar residual plots to those produced in Figure~\ref{fig:model1res}
are shown in Figure~\ref{fig:model2res}, although not perfect, show
a much more reasonable adherence to constancy of variance.

\begin{figure}
\centering
\resizebox{\textwidth}{!}{\includegraphics{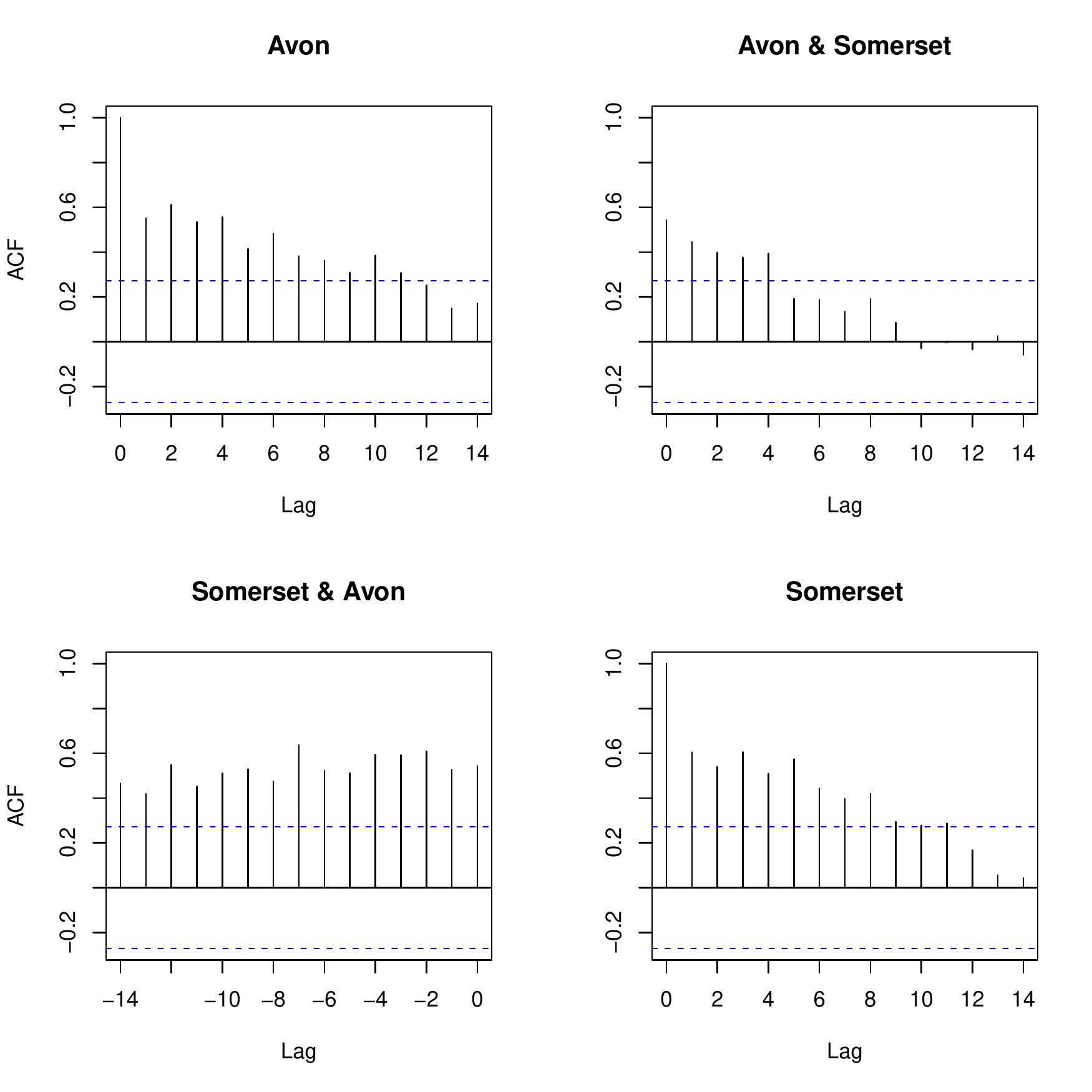}}
\caption{Cross covariance of log population corrected mumps series
for Somerset and Avon (neighbouring counties).\label{fig:acf136}}
\end{figure}

We can further investigate the correlation structure of the residual series
by using a cross-covariance analysis.
Figure~\ref{fig:acf136} shows the cross-covariance analysis
for the $Y_{i, t}$ series (log population corrected mumps).
The autocorrelation plots for Avon and Somerset show significant autocorrelations
which decay slowy, not inconsistent with an autoregressive structure
or {\em possibly} due to trend.
(we will say more on which it might be in Section~\ref{sec:tdsd}).

At this point we should mention that we have only showed autocorrelation
plots for two counties. Similar plots occur for most of the other pairs and
these two counties are fairly representative behaviour for all that comes below.

A
further partial autocorrelation analysis (not shown) suggests and AR$(2)$
structure for Avon and an AR$(1)$ structure for Somerset. This means that
we should investigate maybe a NAR$\{ 2, (s_1, s_2) \}$ model, for some
$s_1, s_2$ neighbour extent for each of the autoregressive components.
Figure~\ref{fig:acf136} also shows significant cross-correlations between the
two series. Investigation of cross-plots for other pairs of cities (even those further
apart) show similar information.

\begin{figure}
\centering
\resizebox{\textwidth}{!}{\includegraphics{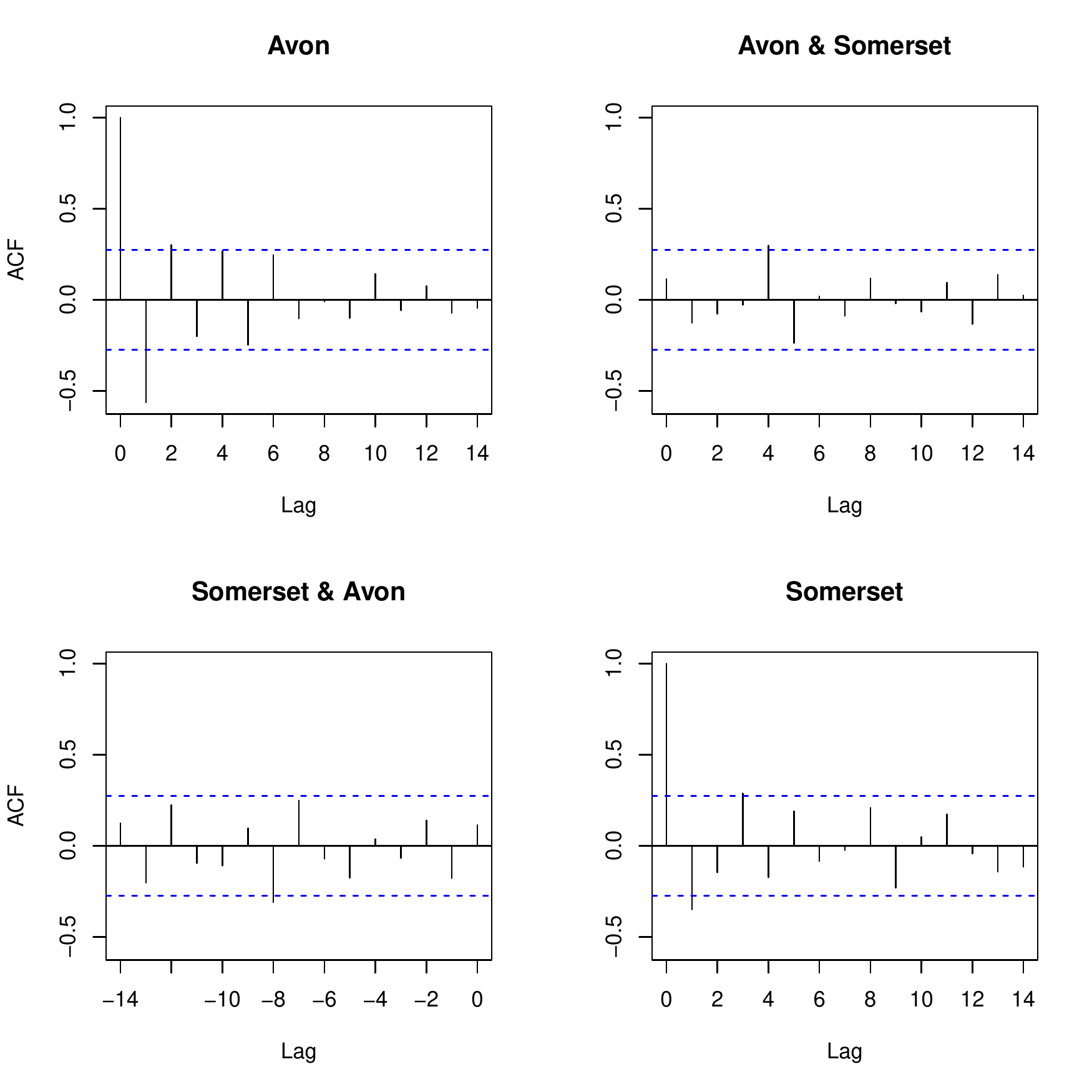}}
\caption{Cross covariance of the residual following the NAR$(1,1)$ fit
for Somerset and Avon (neighbouring counties).\label{fig:acfR136}}
\end{figure}

\begin{figure}
\centering
\resizebox{\textwidth}{!}{\includegraphics{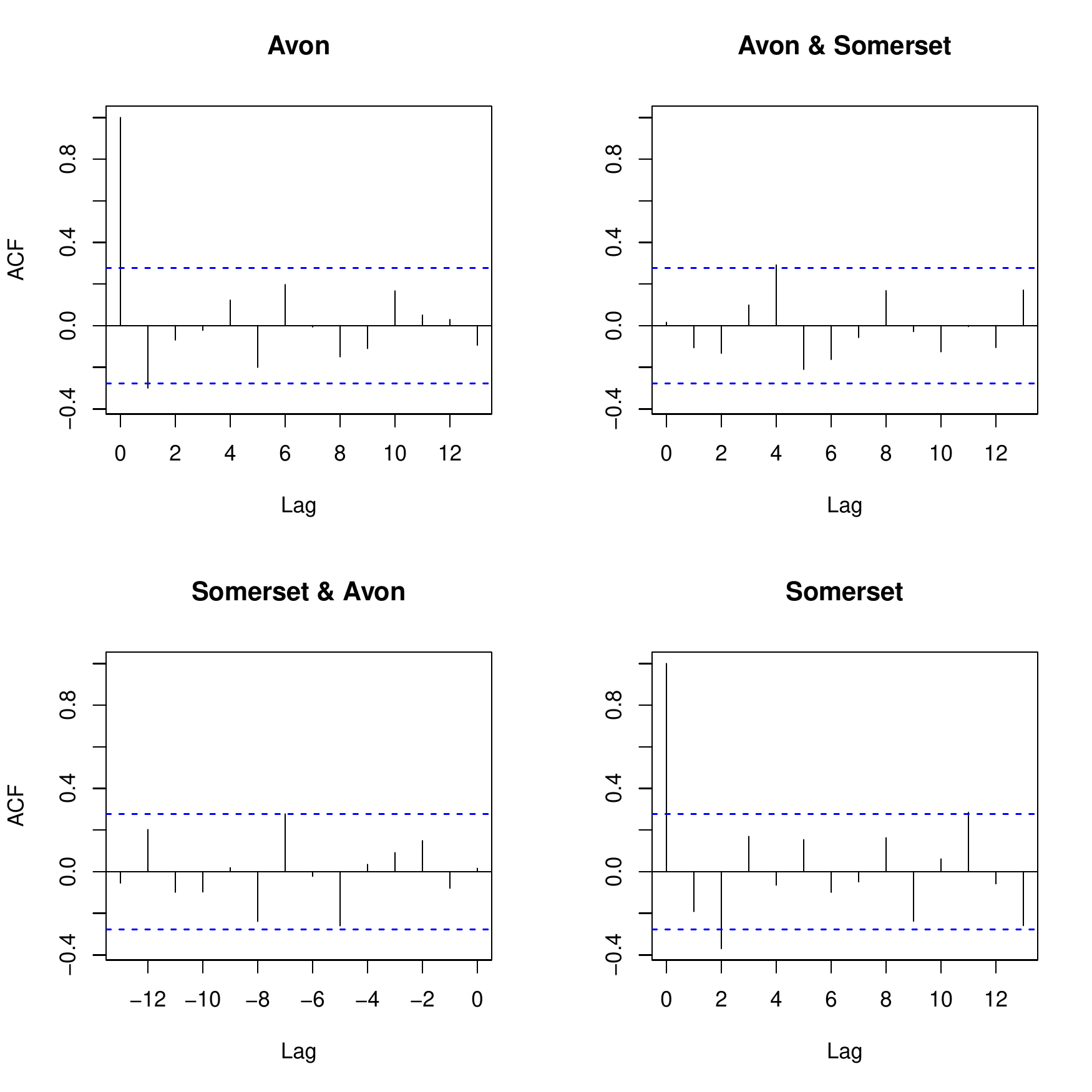}}
\caption{Cross covariance of the residual following the NAR$(2,1)$ fit
for Somerset and Avon (neighbouring counties).\label{fig:acfR136_21}}
\end{figure}

Figure~\ref{fig:acfR136} shows a cross-covariance analysis applied
to the residuals of the NAR$(1,1)$ fit. Very little cross-correlation
exists and much of the correlation in the series themselves is much reduced with
maybe some slight further autoregressive structure remaining to be modelled.
Since we believe there is further autoregressive structure to be modelled
we then fitted a NAR$(2, [1,0])$ model (so, up to lag two standard AR structure, plus
contributions from immediate neighbours of lag-one nodes). The
associated cross-covariance of residuals plot is shown in Figure~\ref{fig:acfR136_21}.
Pleasingly, the autocorrelation plots look more consistent with white noise.
The parameters of the NAR$(2, [1,0])$ model were $\hat{\alpha}_1 \approx 0.394$,
$\hat{\alpha}_2 \approx 0.380$ and $\hat{\beta} \approx 0.204$.

We can also carry out a simple ANOVA analysis shown in Table~\ref{tab:anova}. The table
 shows the benefits of moving from a NAR$(1,0)$ model to a NAR$(2, [1,0])$ model, but
little benefit in going further and using a NAR$(2, [1,1])$ model.
\begin{table}
\centering
\caption{ANOVA of selected NARIMA models on log-transformed data.\label{tab:anova}}
\begin{tabular}{cr}\hline
Model & Residual Sum of Squares\\\hline
NAR$(1,0)$ & 1212.2\\ 
NAR$(1,1)$  & 1029.8\\
NAR$(2, [1,0])$  & 862.2\\
NAR$(2, [1,1])$ & 862.1
\end{tabular}
\end{table}

\section{Trend Removal and Network Differencing}
\label{sec:detrend}
For a network time series there are a large number of possibilities to remove trend.
We propose using the network lifting transform as described
in \cite{JansenNasonSilverman01} and \cite{JNS09}. Effectively, this transform performs
a `wavelet transform on a network' and we do this separately
for each time point on $\{ X_{i, t} \}_{i=1}^K$.
We use the {\tt idnet} function from the {\tt NetTree} {\tt R} package.
Such an operation was first proposed by \cite{NunesKnightNason15} but for the purposes of
decorrelation.

Essentially, the wavelet coefficients act as local spatial differences: indeed, if a node has a single neighbour
then the wavelet coefficient associated with that node is precisely the difference with its neighbour.
As such, our wavelet lifting transform performs a network  operation analogous to
the usual time series differencing $\nabla X_t = X_{t} - X_{t-1}$, but spatially.
\begin{figure}
\begin{minipage}[b]{0.5\linewidth}
\centering
\includegraphics[width=\textwidth]{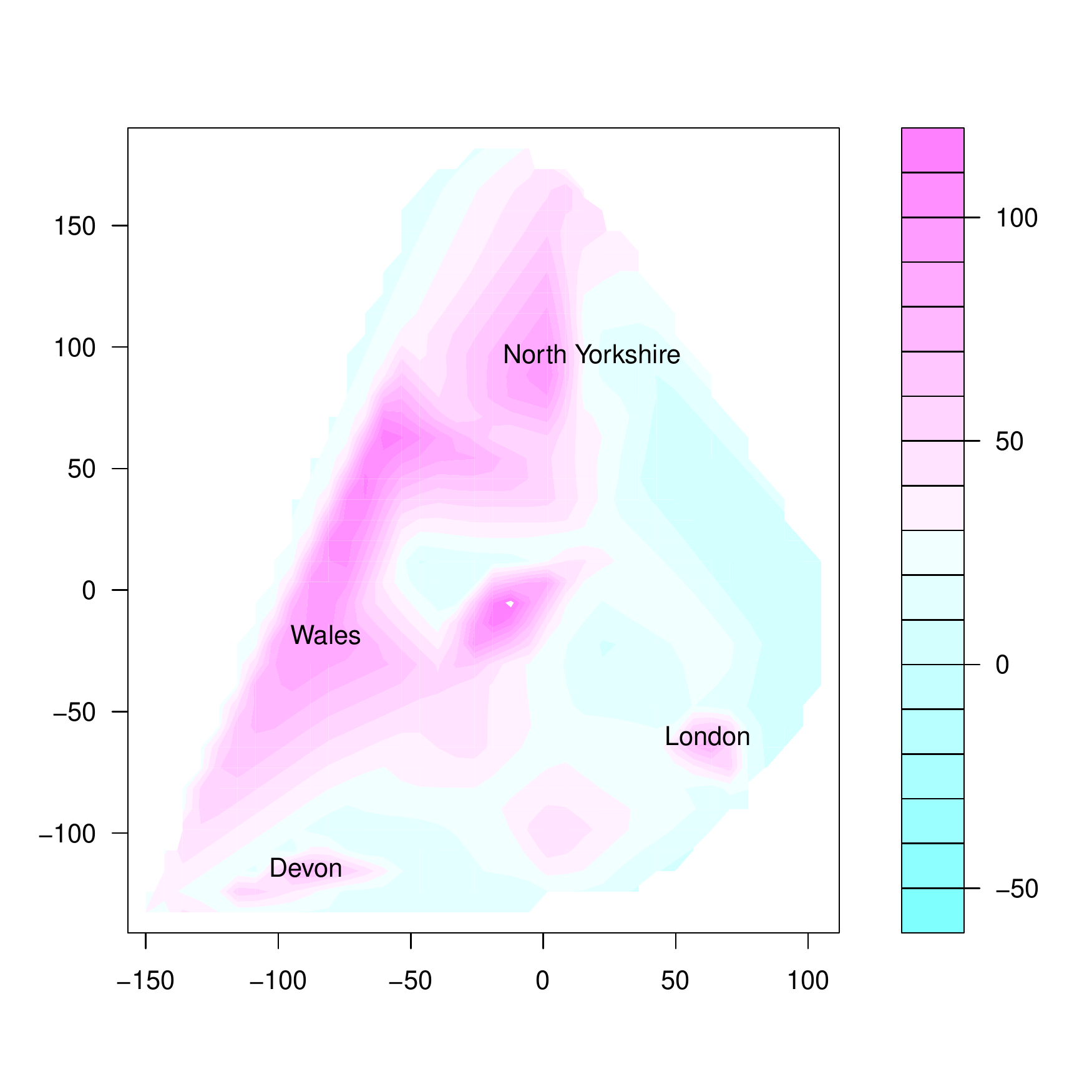}
\end{minipage} \begin{minipage}[b]{0.5\linewidth}
\centering
\includegraphics[width=\textwidth]{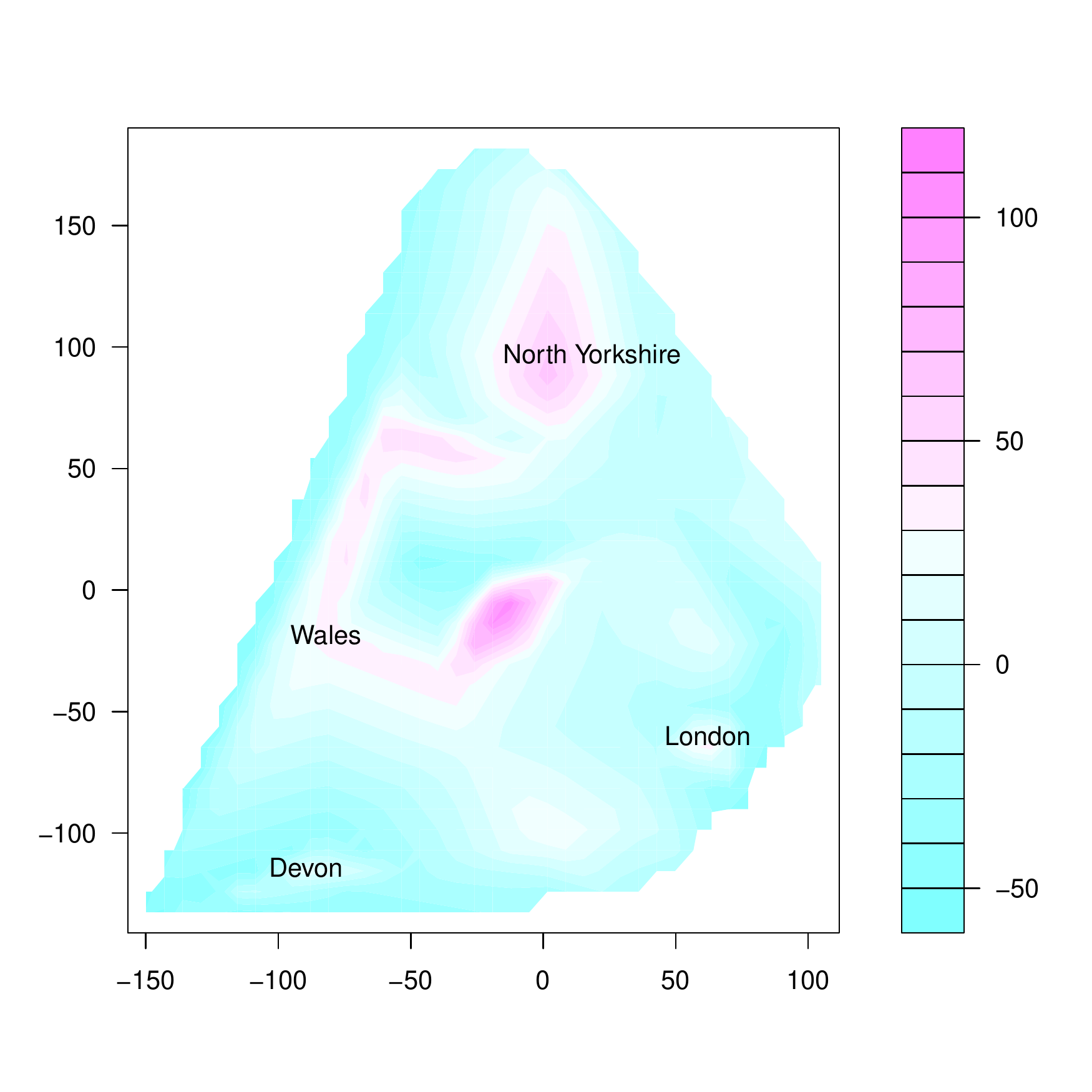}
\end{minipage}
\caption{Left: state of mumps network time series shown at time $t=6$ weeks.
	Right: spatially detrended network at time $t=6$ weeks. Four counties are shown
		for reference. \label{fig:detrend}}
\end{figure}
Figure~\ref{fig:detrend} shows the result of the (spatial) network detrending at time $t=6$ weeks. The
detrended plot shows a much flatter surface with a more constant use of colour
in the detrended plot.
\begin{figure}
\centering
\resizebox{0.7\textwidth}{!}{\includegraphics{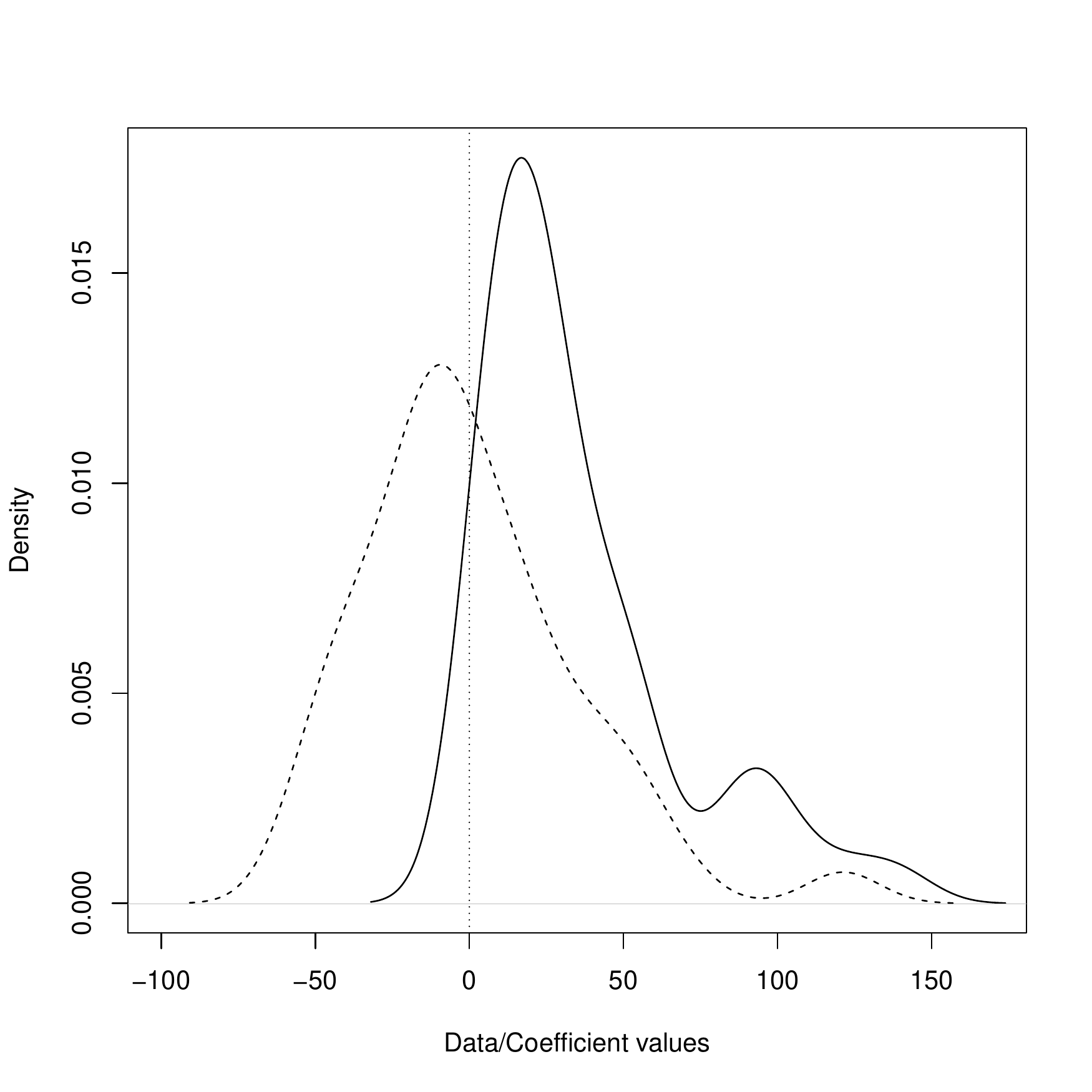}}
\caption{Density of values of mumps cases (solid) and detrended cases (dashed).
	\label{fig:detrenddensity}.}
\end{figure}
The conclusions from this geographical picture is backed up
by the density plots in Figure~\ref{fig:detrenddensity} which shows the detrended values
more tightly packed around zero. Hence, the detrended values are much smaller on the average.

After trend removal we fitted several NARIMA models. Table~\ref{tab:anova2} shows the ANOVA
associated with some of these.
\begin{table}
\centering
\caption{ANOVA of selected NARIMA models on detrended data.\label{tab:anova2}}
\begin{tabular}{cr}\hline
Model & Residual Sum of Squares\\\hline
NAR$(1,0)$ & 1190.309\\ 
NAR$(1,1)$  & 1190.306\\
NAR$(1, [2, 0])$ & 1182.4\\
NAR$[2, [1,0]]$  & 961.1\\
NAR$[2, [1,1]]$ & 959.5
\end{tabular}
\end{table}
Here, we have used a new model the NAR$(1, [2, 0])$ which only uses one-step temporal dependence,
but with neighbours and neighbours of neightbours which corresponds to
\begin{equation}
X_{i, t} = \alpha X_{i, t-1} + \sum_{q \in \N^{(1)}(i)} \beta^{(1)}_q X_{q, t-1} + \sum_{r \in \N^{(2)}(i)} \beta^{(2)}_r X_{r, t-1} + \epsilon_{i, t}.
\end{equation}
Here, the parametrisation of $\beta^{(1)}_q$ is the same as above whereas $\beta^{(2)}_r$ is 
similar but the inverse distance weights are computed from new distances obtained by
adding the distance of `neighbour of neighbour' $r$ to neighbour $q$ and then the distance of
$q$ to the original node $i$. This new total distance represents our best informed guess about the overall
distance from `neighbour of neighbour' $r$ to $i$.

The Table~\ref{tab:anova2} ANOVA shows that neighbours and neighbours of neighbours seem
to have little explanatory power and, at this stage, without further investigation, we might be
better off modelling the series purely as a set of separate univariate autoregressive processes
as proposed and studied in this after-lifting context by \cite{NunesKnightNason15}.

\section{Time-decorrelation achieved by spatial differencing}
\label{sec:tdsd}
An intriguing empirical observation of the study exploited
in \cite{NunesKnightNason15} was that network differencing
(across space) seemed to
result in substantial decorrelation across time and the great practical advantage of
being able to model a network series as a set of separate ARMA processes (i.e.\ not requiring VAR at all).

However, there is an important question. Is the excellent low-autocorrelation
nature of the lifted multivariate series due to trend removal or decorrelation?

To study this phenomenon theoretically we begin with the simplest two-node network and
endow it with a zero mean
two-dimensional VAR(1) time series model. We label the nodes $i$ and $q$. The VAR$(1)$
model we
use parallels the one in~\eqref{eq:nar11} given as follows. Let $\bX_t = (X_{i, t}, X_{q, t})^T$. Then
\begin{equation}
\bX_t = \Pi_1 \bX_{t-1} + \bepsilon_t,
\end{equation}
where 
\begin{equation}
\Pi_1 = \begin{pmatrix}
\alpha & \beta\\
\beta & \alpha
\end{pmatrix},
\end{equation}
and $\bepsilon_t$ is a bivariate white noise process with zero mean and
variance $\sigma^2 I$.
The eigenvalues of $\Pi_1$ are $\alpha+\beta$, $\alpha - \beta$ and
standard texts show that conditions for stationarity are
$| \alpha + \beta | < 1$ and $| \alpha - \beta | < 1$. The region of stationarity can be 
 graphically
 depicted by the square of side length $\sqrt{2}$, centered on the origin rotated by $\pi/4$.
 The stationary covariance matrix of $\bX_t$ is given by
 $\Sigma_X = \sigma^2 \sum_{k=0}^\infty \Pi_1^{2k} = (I - \Pi_1^2)^{-1}$.
Let $\sigma^2_i = \var(X_{i, t})$ and $\sigma^2_{i, q} = \cov( X_{i, t}, X_{q, t})$.
Due to the symmetry in the model  $\sigma^2_q = \sigma^2_i$ for this system.
Moreover, it can be shown that
  the cross-correlation
  $\rho_{i, q} = \sigma_{i, q} / \sigma^2_i = 2\alpha\beta / (1 - \alpha^2 - \beta^2)$.

We desire to study the network
lifted version of this VAR$(1)$ process. The lifted coefficients in this
case are particularly simple and just differences:
\begin{equation}
d_{i, t} = X_{i, t} - X_{q, t}\ \ \ \text{and}  \ \  d_{q, t} = X_{q, t} - X_{i, t} = -d_{i, t},
\end{equation}
for all $t$.

Hence, now we have a lifted VAR$(1)$ process what is its autocorrelation? 
First, we compute the autocovariance of the process $X_{i, t}$ at lag one
\begin{eqnarray}
c_X &=& \cov(X_{i, t}, X_{i, t-1})\\
 &=& \cov( \alpha X_{i, t-1} + \beta X_{q, t-1}, X_{i, t-1} )\nonumber\\
&=& \alpha \sigma^2_i + \beta \sigma_{i, q},
\end{eqnarray}
using $\sigma^2_i =  \sigma^2_q$ from above.

Then, the autocorrelation (at lag one) of the lifted series is
\begin{eqnarray}
c_d &=& \cov( d_{i, t}, d_{i, t-1} )\\
 &=& \cov ( X_{i, t} - X_{q, t}, X_{i, t-1} - X_{q, t-1})\\
&=& \cov \{  \alpha X_{i, t-1} + \beta X_{q, t-1} - ( \alpha X_{q, t-1} + \beta X_{i, t-1} ),
	X_{i, t-1} - X_{q, t-1} \} \nonumber\\
	&=& \cov \{ (\alpha - \beta) X_{i, t-1} + (\beta - \alpha) X_{q, t-1}, X_{i, t-1} - X_{q, t-1} \}\\
	&=& 2 (\alpha - \beta) \sigma^2_i - (\alpha - \beta) \sigma_{i, q}
		+ (\beta - \alpha) \sigma_{q, i}\\
	&=& 2 (\alpha - \beta) ( \sigma^2_i  - \sigma_{i, q}).
\end{eqnarray}
Our empirical results on real data suggested that $|c_d|$ is often less than $|c_X|$ in the general
multivariate lifting situation.
Is this true for our cut-down model?

We can further simplify $c_X, c_d$ by dividing through by $\sigma^2_i$ to obtain
$r_X = \alpha + \beta \rho_{i, q}$ and $r_d = 2(\alpha - \beta)(1 - \rho_{i, q})$ 
where we have an explicit expression for $\rho_{i,q}$ in terms of $(\alpha, \beta)$
from above. We examine:
\begin{equation}
\ad (\alpha, \beta) = |c_X | - |c_d|,
\end{equation}
as the difference of the absolute values of the unlifted and lifted lag-one 
covariance. We are interested in knowing when there is a reduction in
absolute covariance and, hence, ask when is $\ad(\alpha, \beta) > 0$?
\begin{figure}
\centering
\resizebox{0.45\textwidth}{!}{\includegraphics{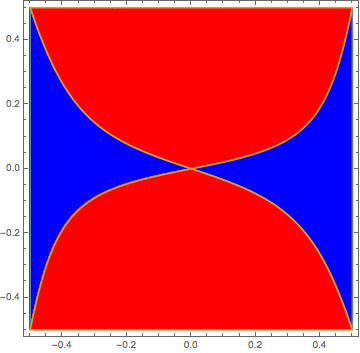}} \hfill
\resizebox{0.45\textwidth}{!}{\includegraphics{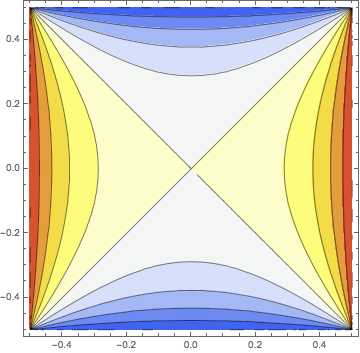}}
\caption{Left: regions where $\ad(\alpha, \beta)<0$ (red) and
	$\ad (\alpha, \beta) > 0$ (blue). Right: contour plot of
	$\rho_{i,q}(\alpha, \beta)$ where deep reds/blues correspond to large
	positive/negative correlations.
	Both plots have  $\alpha - \beta$ on the horizontal
	axis and $\alpha + \beta$ on the vertical axis which correspond to rotating and shrinking the actual stationary region to
	the unit square centred at the origin.\label{fig:ad}}
\end{figure}
Figure~\ref{fig:ad} (left) shows that $\ad (\alpha, \beta)$ is more often negative than
positive, corresponding to an increase in the absolute value of the lag-one
autocorrelation, the opposite of what we might have hoped from \cite{NunesKnightNason15}.
It is also helpful to refer to Figure~\ref{fig:ad} (right) which shows the contours 
of $\rho_{i, q} (\alpha, \beta)$ in the same coordinate system as the left plot.
It can be seen that $|\rho_{i, q}|$ is large near the boundaries of the region and
from the left plot it can be seen the negative spatial correlations are associated with an increase
in absolute autocorrelation, but with positive spatial correlations are associated with a   decrease.
\begin{figure}
\centering
\resizebox{\textwidth}{!}{\includegraphics{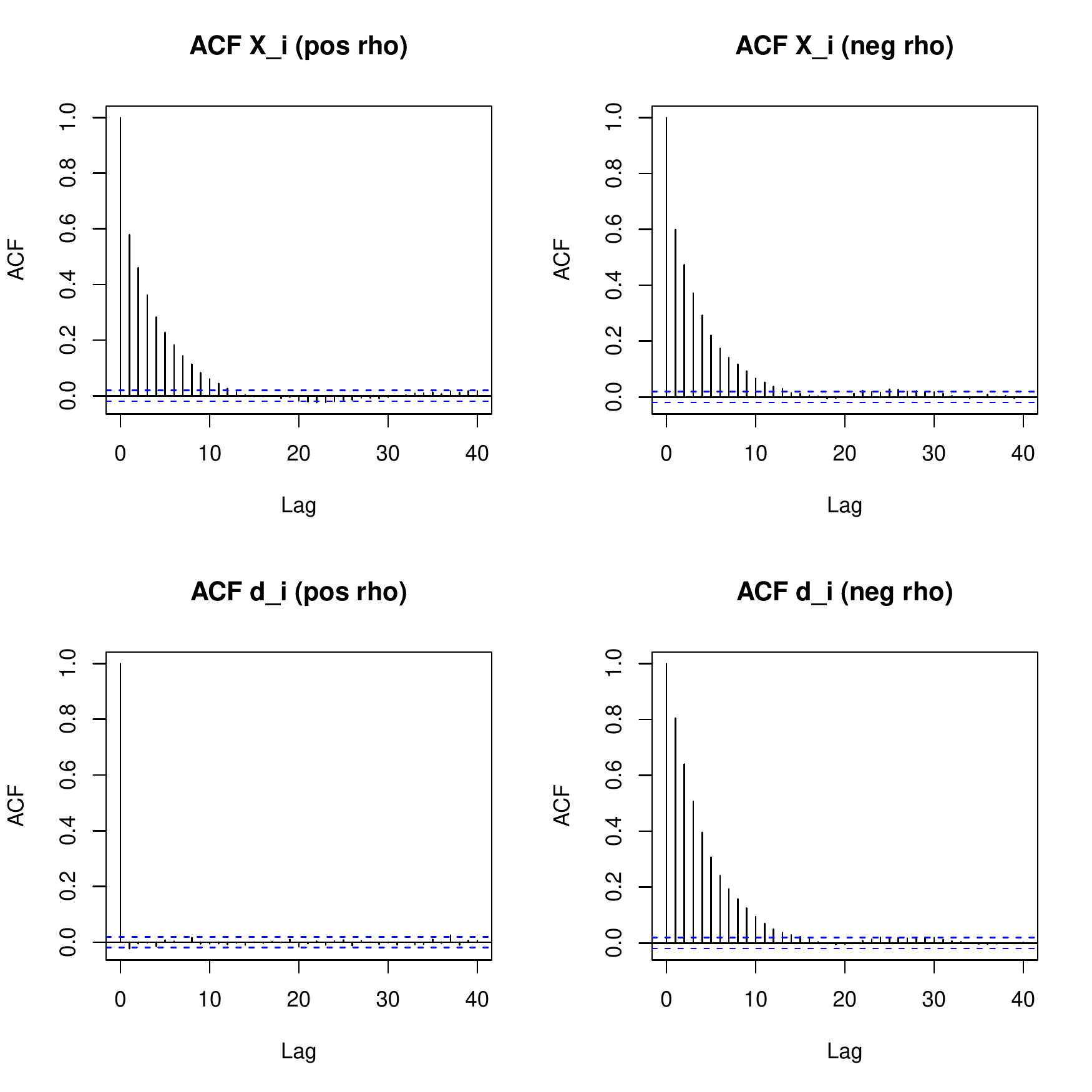}}
\caption{Autocorrelation plots of the $X_{i, t}$ series (top row) and
$d_{i, t}$ series (bottom row) for the situation $\alpha = \beta = 0.4$ (left column)
and $\alpha = 0.4, \beta = -0.4$ (right column).\label{fig:fouracf}}
\end{figure}
Figure~\ref{fig:fouracf} shows the above result in action.
For $\alpha = \beta = 0.4$ the quantity $\rho_{i, q} = 8/17$ is positive and, according
to the theory, should result in a decrease in the absolute autocorrelation.
Indeed, comparing the bottom with top plot in the left column of Figure~\ref{fig:fouracf}
shows that  the autocorrelations at lag one (and the rest) are all much smaller.
However, for $\alpha = 0.4, \beta = -0.4$ we have $\rho_{i, q} = -8/17$ the theory
says that the autocorrelation should increase in absolute value and, indeed,
looking at the right column from top to bottom in Figure~\ref{fig:fouracf} this
is indeed the case.

Overall, the message is that there is not a uniform reduction in the absolute
value of autocorrelation, nor is it possible to say that it mostly happens.
Whether it happens or not depends specifically on the choice of the parameters
in this model, 

At this point it should be stressed that this is a very simple case. The theory
above only examines a simple bivariate VAR model on a network with two
nodes. Our mumps network has 47 nodes and many real networks are much
larger. For the two node network the lifting step is particularly simple (differencing
of neighbouring values). For larger networks, nodes with a larger number of
neighbours are not merely subject to differencing but subtract off
some linear combination of its neighbour values. We conjecture when these are positively
correlated then there will be a reduction in correlation, but for those situations
where negative correlations are involved, the result will be unpredictable.
Further study is required in this area.

In practical terms, it seems  likely that our network differencing, $\Dfrak$,
is responsible for removing trend which is causing slow autocorrelation decay akin
to integrated processes in classical time series analysis.

\section*{Acknowledgements}
We would like to thank Douglas Harding and Daniela DeAngelis of the Health Protection Agency
for supplying the Mumps data.

\end{document}